\documentclass[onecolumn]{aastex63}
\usepackage{xspace}

\newcommand\aastex{AAS\TeX}

\newcommand{\notes}[1]{}
\newcommand{\comment}[1]{}

\newcommand{\muas}{{$\mu$as}\xspace}

\def\apjl{\rm{ApJ}}

\newcommand{\ehtim}{\texttt{eht-imaging}\xspace}

\newcommand{\JL}[1]{\textcolor{black}{#1}}

\defcitealias{EHT1}{EHTC~I}
\defcitealias{EHT2}{EHTC~II}
\defcitealias{EHT3}{EHTC~III}
\defcitealias{EHT4}{EHTC~IV}
\defcitealias{EHT5}{EHTC~V}
\defcitealias{EHT6}{EHTC~VI}
\defcitealias{EHT7}{EHTC~VII}
\defcitealias{EHT8}{EHTC~VIII}

\submitjournal{ApJ}

\shorttitle{VLBI-NET ApJ}
\shortauthors{Lin et al.}

\graphicspath{{./}{figures/}}

\begin{document}

\title{Template \aastex Article with Examples: 
v6.3\footnote{Released on June, 10th, 2019}}

\title{VLBInet: Radio Interferometry Data Classification for EHT with Neural Networks}

\correspondingauthor{Joshua Y.-Y. Lin}
\email{yaoyuyl2@illinois.edu}

\author[0000-0003-0680-4838]{Joshua Y.-Y. Lin}
\affiliation{Department of Physics, University of Illinois, 1110 West Green Street, Urbana, IL 61801, USA}
\affiliation{Illinois Center for Advanced Study of the Universe, 1110 West Green Street, Urbana, IL 61801, USA}
\affiliation{National Center for Supercomputing Applications, 605 East Springfield Avenue, Champaign, IL 61820, USA}
\affiliation{Center for Computational Astrophysics, Flatiron Institute, Simons Foundation, 162 5th Ave, New York, NY 10010, USA}

\author[0000-0002-5278-9221]{Dominic W. Pesce}
\affiliation{Center for Astrophysics $\; \vert \;$ Harvard $\,\&\,$ Smithsonian, 60 Garden Street, Cambridge, MA 02138, USA}
\affiliation{Black Hole Initiative at Harvard University, 20 Garden Street, Cambridge, MA 02138, USA}

\author[0000-0001-6952-2147]{George N. Wong}
\affiliation{Department of Physics, University of Illinois, 1110 West Green Street, Urbana, IL 61801, USA}
\affiliation{Illinois Center for Advanced Study of the Universe, 1110 West Green Street, Urbana, IL 61801, USA}
\affiliation{School of Natural Sciences, Institute for Advanced Study, 1 Einstein Drive, Princeton, NJ 08540, USA}
\affiliation{Princeton Gravity Initiative, Princeton University, Princeton, New Jersey 08544, USA}

\author[0000-0002-8031-9323]{Ajay Uppili Arasanipalai}
\affiliation{Department of Physics, University of Illinois, 1110 West Green Street, Urbana, IL 61801, USA}

\author[0000-0002-0393-7734]{Ben~S.~Prather}
\affiliation{Department of Physics, University of Illinois, 1110 West Green Street, Urbana, IL 61801, USA}
\affiliation{Illinois Center for Advanced Study of the Universe, 1110 West Green Street, Urbana, IL 61801, USA}

\author[0000-0001-7451-8935]{Charles F. Gammie}
\affiliation{Department of Astronomy, University of Illinois, 1002 West Green Street, Urbana, IL 61801, USA}
\affiliation{Illinois Center for Advanced Study of the Universe, 1110 West Green Street, Urbana, IL 61801, USA}
\affiliation{National Center for Supercomputing Applications, 605 East Springfield Avenue, Champaign, IL 61820, USA}
\affiliation{Department of Physics, University of Illinois, 1110 West Green Street, Urbana, IL 61801, USA}

\begin{abstract}

The Event Horizon Telescope (EHT) recently released the first horizon-scale images of the black hole in M87.  Combined with other astronomical data, these images constrain the mass and spin of the hole as well as the accretion rate and magnetic flux trapped on the hole.  An important question for the EHT is how well key parameters, such as trapped magnetic flux and the associated disk models, can be extracted from present and future EHT VLBI data products. The process of modeling visibilities and analyzing them is complicated by the fact that the data are sparsely sampled in the Fourier domain while most of the theory/simulation is constructed in the image domain. Here we propose a data-driven approach to analyze complex visibilities and closure quantities for radio interferometric data with neural networks. Using mock interferometric data, we show that our neural networks are able to infer the accretion state as either high magnetic flux (MAD) or low magnetic flux (SANE), suggesting that it is possible to perform parameter extraction directly in the visibility domain without image reconstruction. We have applied VLBInet to real M87 EHT data taken on four different days in 2017 (April 5, 6, 10, 11), and our neural networks give a score prediction $0.52, 0.4, 0.43, 0.76$ for each day, with an average score $0.53$, which shows no significant indication for the data to lean toward either the MAD or SANE state.

\end{abstract}

\section{Introduction}

The Event Horizon Telescope (EHT) is a globe-spanning network of millimeter wavelength observatories~\citepalias{EHT1, EHT2}.  Data from individual observatories are combined to measure the Fourier components of a source {\em intensity} on the sky~\citepalias{EHT3}.  The sparse set of Fourier components together with a regularization procedure can then be used to reconstruct an image of the source~\citepalias{EHT4}.  The resulting images of the black hole at the center of M87 (hereafter M87*) have a ringlike structure---attributed to emission from hot plasma surrounding the black hole---with an asymmetry that contains information about the motion of plasma around the hole~\citepalias{EHT5, EHT6}.  Combined with data from other sources, the EHT images constrain the black hole spin and mass as well as properties of the surrounding magnetic field structure and strength~\citepalias{EHT5, EHT6}. More recently, the EHT has produced polarized emission maps of M87* \citepalias{EHT7}, and these imply a highly organized magnetic field structure in the source~\citepalias{EHT8}.

Black hole accretion flows can be divided into two qualitatively different states, called MAD and SANE, depending on the strength of the magnetic field near the event horizon. In the magnetically arrested disk (MAD) state, magnetic fields near the horizon are dynamically important enough to limit the flow of plasma onto the hole \citep{Kogan1974,ichimaru1977mad,igumenschchev2003mad,narayan2003mad}. When MAD, the magnetic flux $\Phi_B$ through the black hole horizon is at the maximum value sustainable by the accretion flow; additional flux escapes outward through the inflowing plasma.  The flux can be  characterized by the dimensionless flux $\phi = \Phi_B / \sqrt{\dot{M} R_g^2 \, c}$ ($\dot{M} \equiv$ accretion rate; $R_g \equiv G M/c^2$).  In the MAD state $\phi \simeq 15 $.\footnote{We use Lorentz--Heaviside units rather than Gaussian units to be consistent with \citealt{Porth2019}. In the latter system, $\phi_c \approx 50$.}   In the standard and normal evolution (SANE) state, by contrast, $\phi < \phi_c$ and the flow is organized in a conventional, geometrically thick, centrifugally supported disk.  For a fixed accretion rate and black hole spin, MAD flows have stronger jets driven by the Blandford--Znajek effect. Is it possible to identify whether a source is in the MAD or SANE state directly from VLBI data?

Various methods have been proposed for using black hole images or VLBI data to constrain system parameters---see \citep{johnson2020universal, gralla2020observable, lupsasca2018critical} for measuring black hole parameters using long-baseline visibilities and \citep{wong2021glimmer, hadar2020photon, chesler2021light} through autocorrelations of position-dependent light echoes---however, 
the general parameter extraction problem is challenging because simulated images contain an abundance of information about not only the black hole spacetime and gross accretion flow properties, but also about fluctuating, turbulent structure in the accreting plasma.  Past efforts to compare simulated GRMHD images with VLBI data have typically either fit to the visibilities constructed from simulated images (e.g., the ``average image scoring'' procedure from \citetalias{EHT5}, which runs into issues of attempting to match or marginalize over specific instantiations of turbulence between simulation and observation) or else have focused on observationally accessible proxies for physical quantities (e.g., the metric derived in \citealt{palumbo2020beta2} and used in \citetalias{EHT8}).  Here we consider a neural network (NN)-based approach to parameter extraction from VLBI data.  This approach has the benefit of using all the data (as opposed to observationally accessible proxies) and can ideally distill from the data all information that constrains a model parameter.

Earlier work by~\citet{van2020deep} found evidence that neural networks could be used to infer values for a limited set of parameters in a library of synthetic black hole images of SANE flows. Work by \cite{lin2020feature} explored both MAD and SANE models and investigated neural network interpretability. Both these earlier efforts worked in the image domain, though the real data correspond to sparse sampling in the Fourier domain. In recent years, work by \cite{sun2020deep, sun2020learning, morningstar2018analyzing} have shown that deep neural networks can be used for visibility-to-image reconstruction, as well as tracking their underlying distribution in a probabilistic way using certain evaluation loss metrics. Work by \citet{popov2021proof} used convolutional neural networks to estimate an inclination angle from complex visibility with synthetic disk-based model. Here we explore the possibility of using neural networks to perform an end-to-end analysis of synthetic M87* data in the visibility domain. 

Our paper is organized as follows. In Section \ref{sec: synthetic_data_generation}, we describe our procedure for generation of synthetic observations. In Section \ref{sec: nn} we describe our neural network, and we present results in Section \ref{sec: results} and provide a discussion in Section \ref{sec: discussion}.

\section{Synthetic data generation}
\label{sec: synthetic_data_generation}

\begin{figure*}[t]
\vskip 0.2in
\begin{center}
\includegraphics[width=\linewidth]{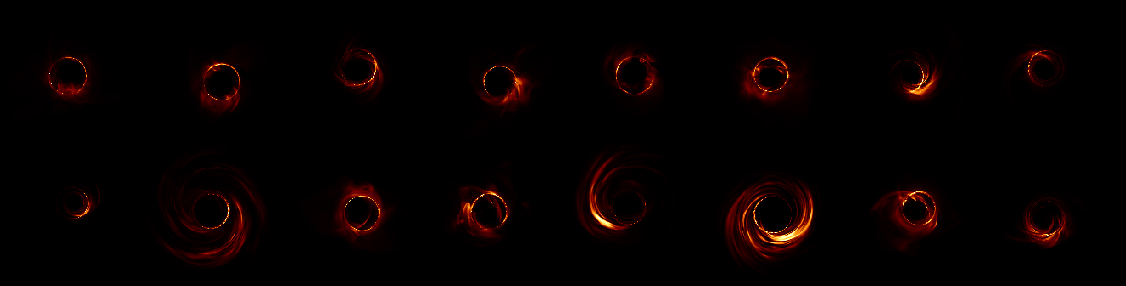}
\caption{Example synthetic images based on numerical models of black hole accretion flows with M87-like parameters.  Top row: sample images of MAD accretion flows. Bottom row: sample images of SANE flows.
}
\label{fig:Synthetic-black-hole-image}
\end{center}
\vskip -0.2in
\end{figure*}

\subsection{Simulating black hole images with GRMHD/GRRT}

The training data 
for our supervised machine learning pipeline 
consist of synthetic images based on general relativistic magnetohydrodynamic (GRMHD) simulations of black hole accretion.
The image dataset used in this work is the same as the one used in \citep{lin2020feature}. We obtained $40000$ images from the image library, and split them into half for training and testing.

The GRMHD simulations were generated with the {\tt{}iharm3D} code (Prather et al.~JOSS, submitted) for five black hole spins ($a_* = -0.94, -0.5, 0, 0.5, 0.94$) in both MAD and SANE states. Notice that there is a continuum of possible SANE states; our SANE models have $\phi \simeq 1$.  For selected spins and magnetic states we consider multiple GRMHD simulations initialized with different seed perturbations to assess the likelihood that our neural networks are over-fitting the simulations.

We then imaged the GRMHD simulation output using  
{\tt ipole} ~\citep{moscibrodzka2018}. To do this we must specify additional physical parameters: the approximate size of the black hole and its distance from the observer; the accretion rate (set so that the intensity integrated over the images matches the mm flux density of M87*); and a parameter $R_{\mathrm{high}}$ that regulates assignment of the electron temperature from fluid variables (see \citetalias{EHT5}).  $R_{\mathrm{high}}$ is drawn uniformly from a discrete set of allowed values $[1, 10, 40, 160]$.  In addition we must specify parameters related to the imaging process: the position angle (PA) of the source, the angular resolution of the image, and the coordinates of the black hole center on the image.  To prevent the neural network from correlating imaging parameters with source parameters we introduce image-by-image variations: the PA is drawn uniformly from $[0, 2\pi)$; the angular scale is drawn uniformly from a small interval corresponding to a $\pm 10\%$ change in angular scale of the source (equivalently a 10\% range in black hole mass or distance); the image center is drawn uniformly from a square region corresponding to a $[-10,10]$ \muas displacement in each coordinate.  

\begin{figure*}[t]
\vskip 0.2in
\begin{center}
\includegraphics[trim = 2cm 0cm 2cm 0mm,
clip,width=\linewidth]{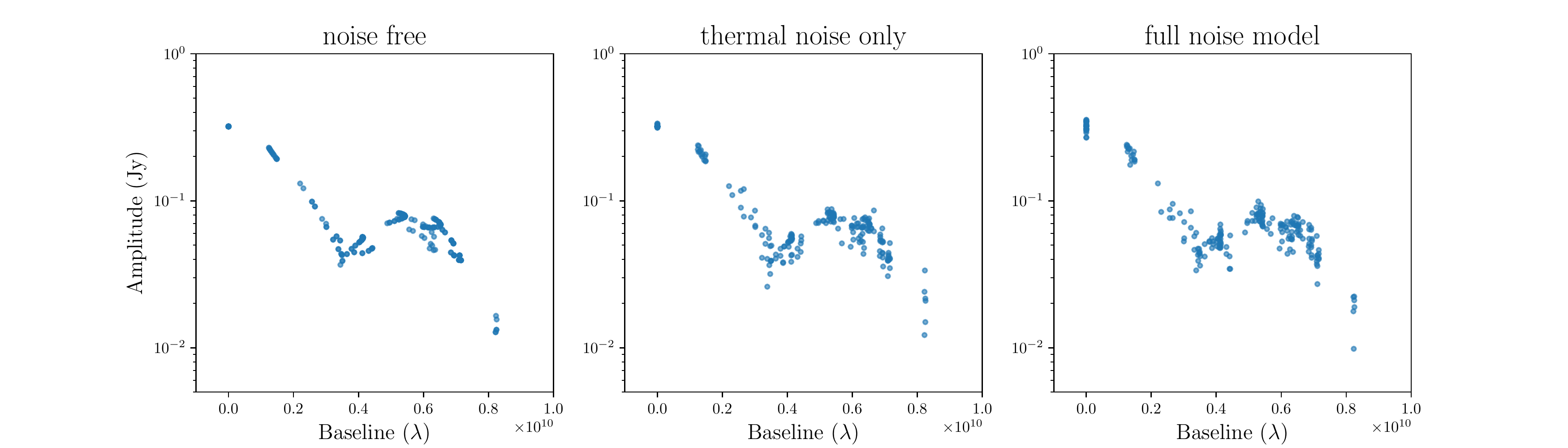}
\caption{Visibility amplitude versus baseline length for a \JL{randomly-selected single training image} with different noise components applied.  Panel (a) shows the noise-free Fourier samples, panel (b) shows these samples after the addition of complex Gaussian noise, and panel (c) shows the same as (b) after additionally modulating each visibility by station-based complex gain noise (see \autoref{sec:VisGen})
}
\label{fig:visibility_panel}
\end{center}
\vskip -0.2in
\end{figure*}

\subsection{Synthetic Data Generation} \label{sec:VisGen}

For each image in the training and testing sets, we generated synthetic VLBI observations using \ehtim  \citep{Chael_2016,Chael_2018} in a manner similar to that described in \citetalias{EHT4}.  The baseline coverage and observing cadence match the 2017 April 11 observations of M87 carried out by the EHT after coherently time-averaging each baseline's complex visibilities on a per-scan basis \citepalias{EHT2,EHT3}.  The source image in each case is assumed to be located at the sky position of the M87 black hole, and the baseline thermal noise values have been taken from the original calibrated observations provided by the EHT collaboration \citep{EHT_data}.




In the absence of noise the complex visibilities are, according to the van Cittert--Zernike theorem \citep{TMS}, samples of the Fourier transform of the sky emission.
Real-world telescope arrays suffer from a number of a priori unknown systematics, however, the most severe of which are variations in the phase and amplitude of the signal received at each station caused by atmospheric turbulence and other signal path effects.  

We have thus generated three classes of mock observations for each of the training and testing images: (1) a ``noise free'' observation that contains only the Fourier transform of the input image sampled on the EHT baselines, (2) a ``thermal noise'' observation in which each complex visibility has been modulated according to its measurement uncertainty, and (3) a ``full noise'' observation that includes station-based complex gain fluctuations.  The synthetic gain amplitudes are modeled as Gaussian-distributed with unit mean and a standard deviation of 10\% for all stations, while the synthetic gain phases are drawn uniformly from $[0, 2\pi)$; both gain amplitudes and phases are sampled independently for each station at each timestamp.

Gain variations are particularly severe at the short ($\sim$1.3 mm) observing wavelength of the EHT, motivating the construction and use of so-called ``closure'' data products that are immune to station-based corruption.  The two standard sets of closure quantities are closure phases and closure amplitudes, which are constructed using triangles and quadrangles of baselines, respectively.  Denoting a visibility measurement on the baseline between stations $i$ and $j$ as $V_{ij}$, the closure phase on triangle $ijk$ is given by the argument of the directed triple product,
\begin{equation}
\psi_{ijk} = {\rm arg}\left( V_{ij} V_{jk} V_{ki} \right) .
\end{equation}
The closure amplitude on a quadrangle $ijk\ell$ is given by
\begin{equation}
A_{ijk\ell} = \left| \frac{V_{ij} V_{kl}}{V_{i\ell} V_{jk}} \right| .
\end{equation}
Both closure phases and closure amplitudes have the property that station-based noise cancels, making them robust observables.  In this paper we construct two classes of neural networks, one to treat complex visibilities and the other to treat closure quantities.



\begin{figure*}
 \centering
  \includegraphics[width=140mm]{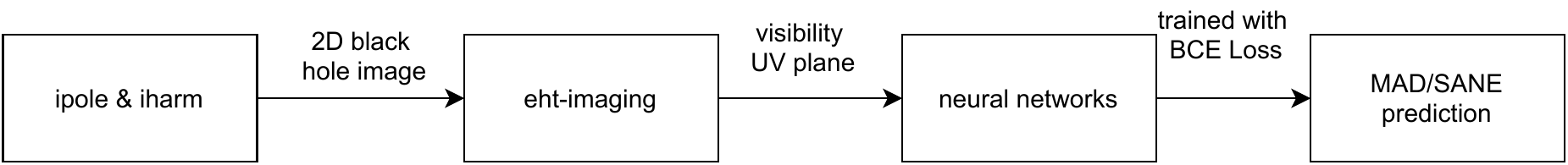}
    \caption{VLBInet Pipeline.}
  \label{fig:pipeline}
\end{figure*}

\section{Neural networks}
\label{sec: nn}

\begin{figure*}
\centering
\begin{tabular}{ccc}
\includegraphics[width=0.46\linewidth]{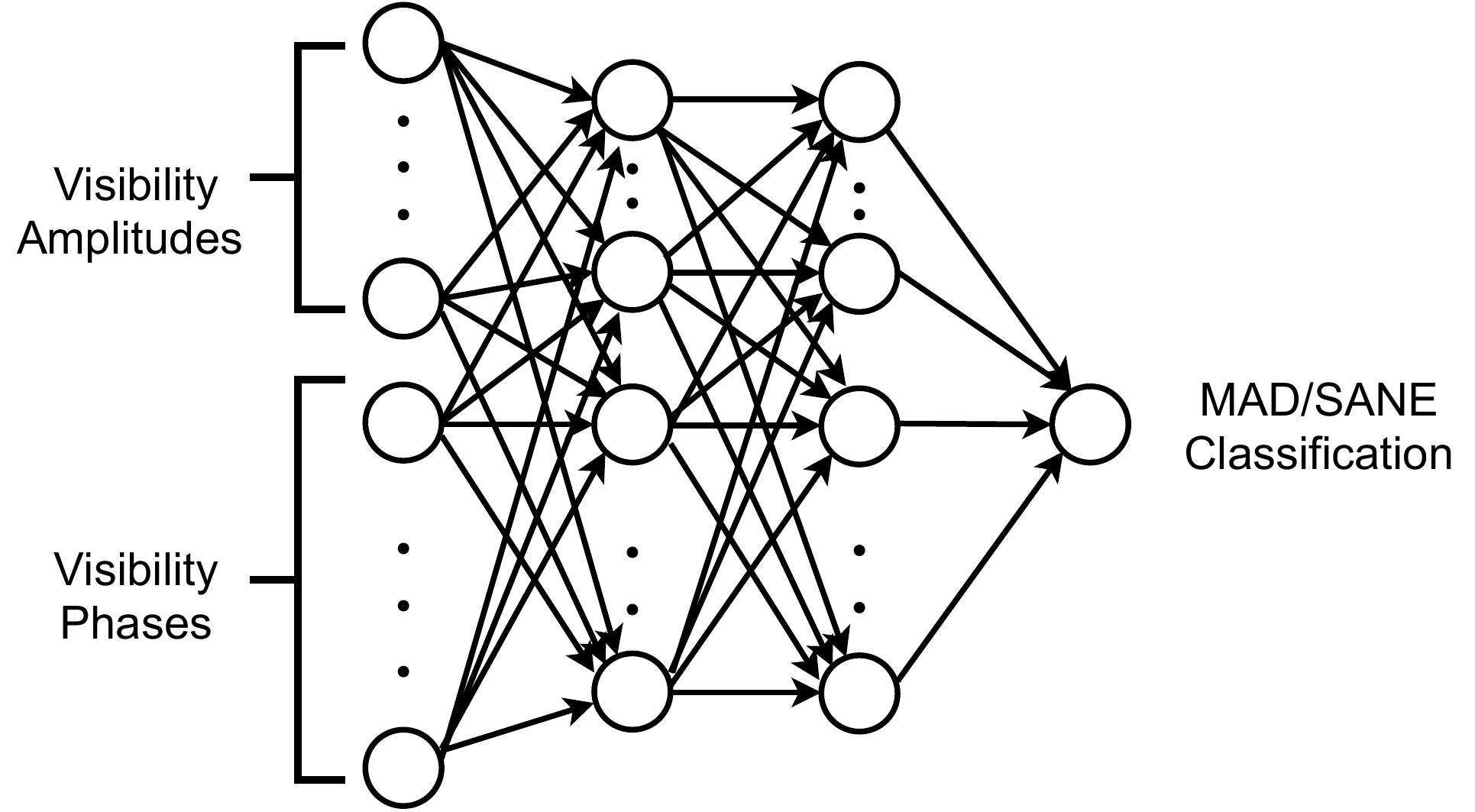}&
\includegraphics[width=0.46\linewidth]{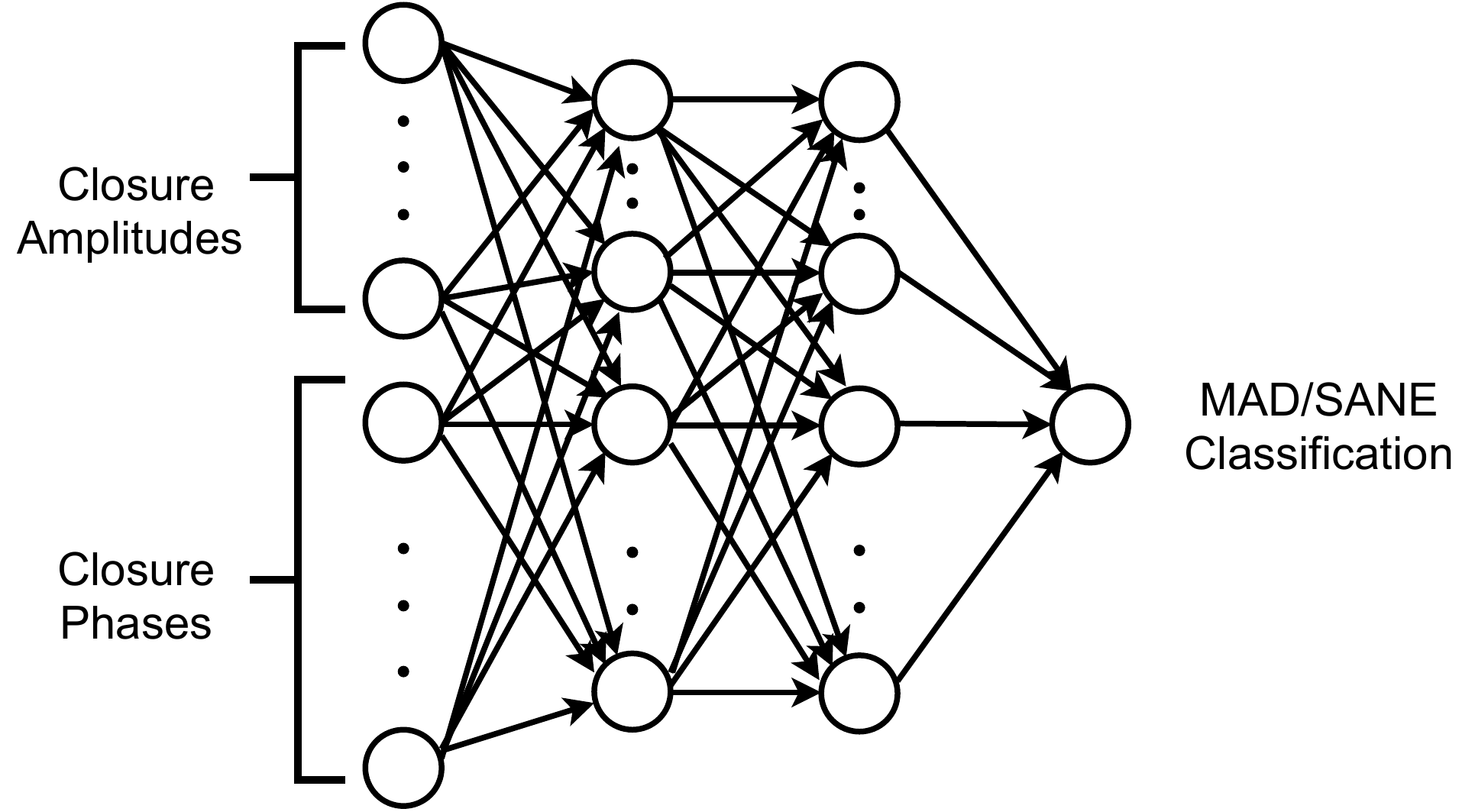}&
\end{tabular}
\caption{Two neural network classes used in this paper. The first accepts complex visibilities as input, and the second accepts closure quantities as input. }
\label{fig:vd_map}
\end{figure*}

We decided to use fully connected neural networks, also known as multilayer perceptrons (MLPs), for the MAD/SANE classification task.  Previous work used convolutional NNs (CNNs) and image domain data \citep{van2020deep, lin2020feature}.  We use sparsely sampled visibility domain data---where the input is not translation invariant---and therefore selected the MLP architecture as one of the simplest architectures that avoids possible inductive biases associated with the translational invariance embedded in, for example, CNNs \citep{battaglia2018relational}.  We will refer to this arrangement as VLBInet.

When working with complex visibilities, we decompose the complex numbers into amplitudes and phases and then project the phases into sines and cosines to enable phase wrapping. For both complex visibilities and closure quantities the data are organized in a consistent way so that each neuron in the input layer is associated with a particular point in the uv domain (complex visibilities) or triangle or quadrangle (closure quantities).  

We trained 6 different versions of VLBInet, ranging over 2 different visibility inputs (complex/closure) and 3 different noise scenarios. All versions consist of 7 layers. For complex visibilities, the numbers of neurons at each layer are $648, 256, 64, 64, 64, 64, 1$ and the full model has $195,137$ trainable parameters. For closure quantities the numbers of neurons at each layer are $381, 256, 64, 64, 64, 64, 1$ and the model has $126,785$ trainable parameters.  In MLPs each connection contains a weight and a bias.  We use $\tanh$ as the activation function.  We use binary cross-entropy (BCE) loss for the MAD/SANE since the output is expected to be in one of these two categories: $\text{BCE Loss} = t \log(p) + (1-t) \log(1 - p)$, where $t = 0$ for MAD and $t = 1$ for SANE.  Here p is the score prediction with range from $0-1$.  VLBInet is built using the Python 3 deep learning library \emph{pytorch} \citep{NEURIPS2019_9015}. 

\section{Results}
\label{sec: results}

\subsection{Results on synthetic data}

\begin{figure*}
\centering
\begin{tabular}{ccc}
\includegraphics[width=0.5\linewidth]{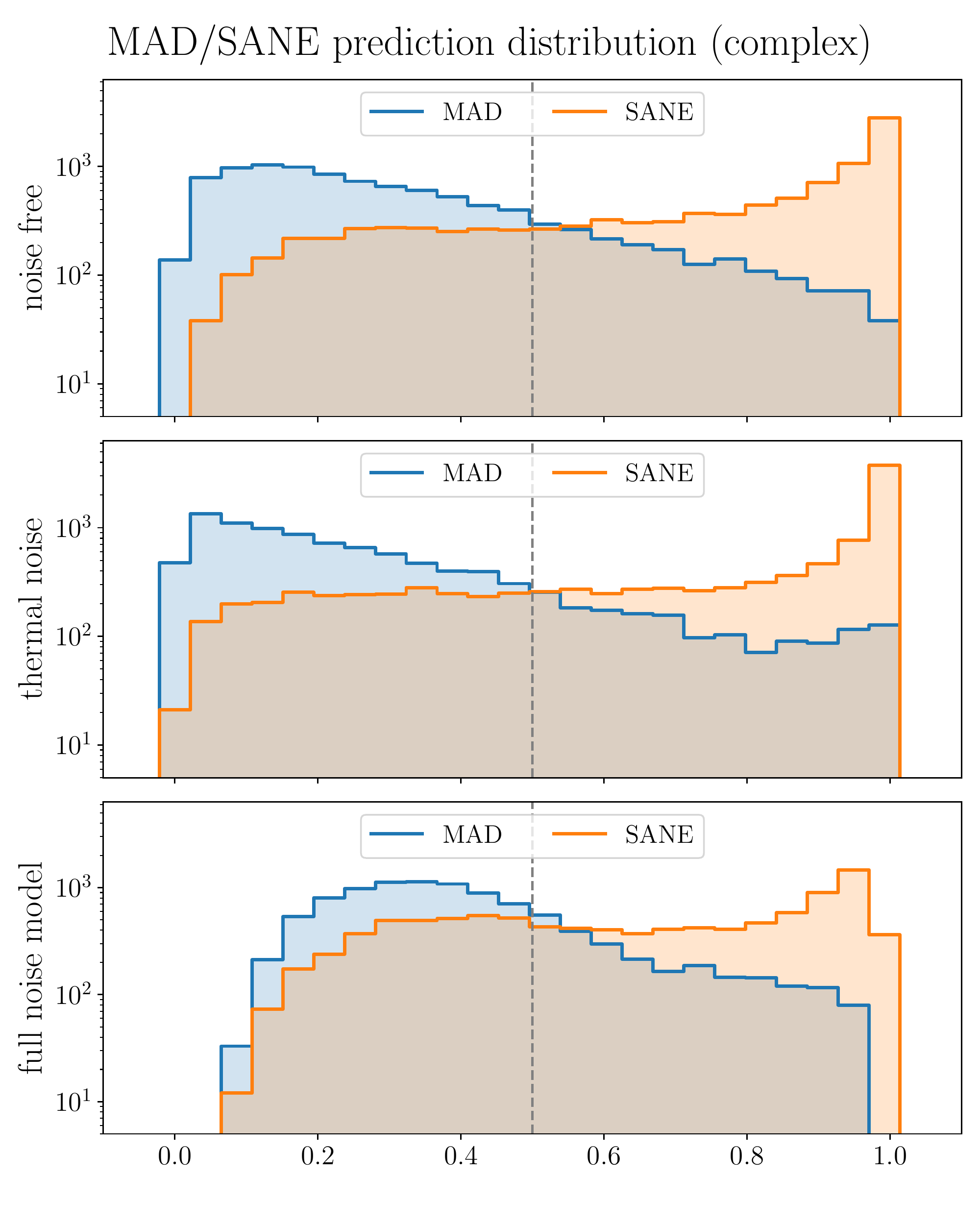}&
\includegraphics[width=0.5\linewidth]{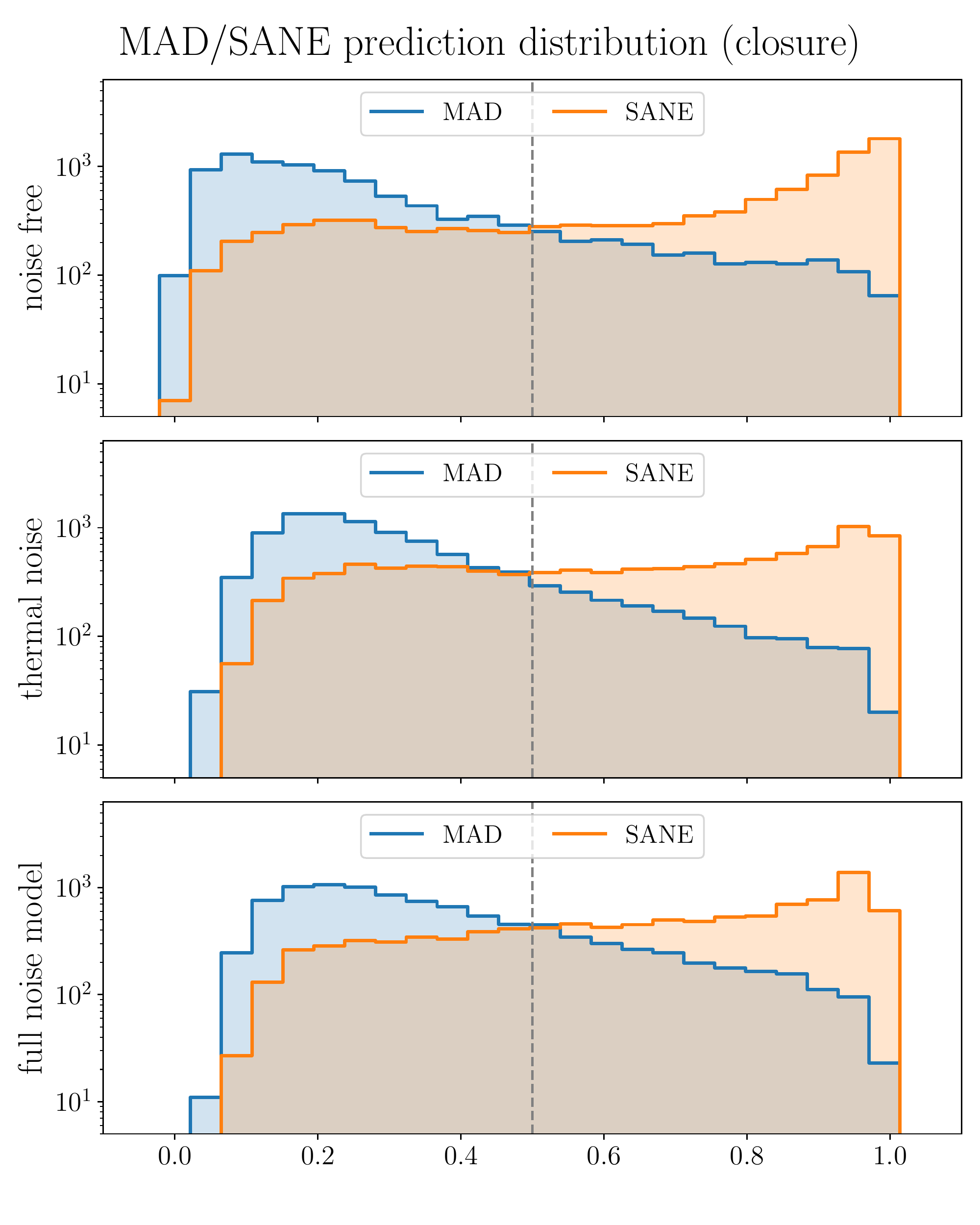}&
\end{tabular}
\caption{
Histogram of scores for MAD and SANE models.  If the neural network could identify the model perfectly then all MADs would score 0 and all SANEs would score 1.   The grey dashed line is the decision boundary: data with score $>$ 0.5 are classified as SANE model, and $<$ 0.5 are classified as MAD model.
}
\label{fig:histogram}
\end{figure*}

\begin{figure*}
\centering
\begin{tabular}{ccc}
\includegraphics[width=0.5\linewidth]{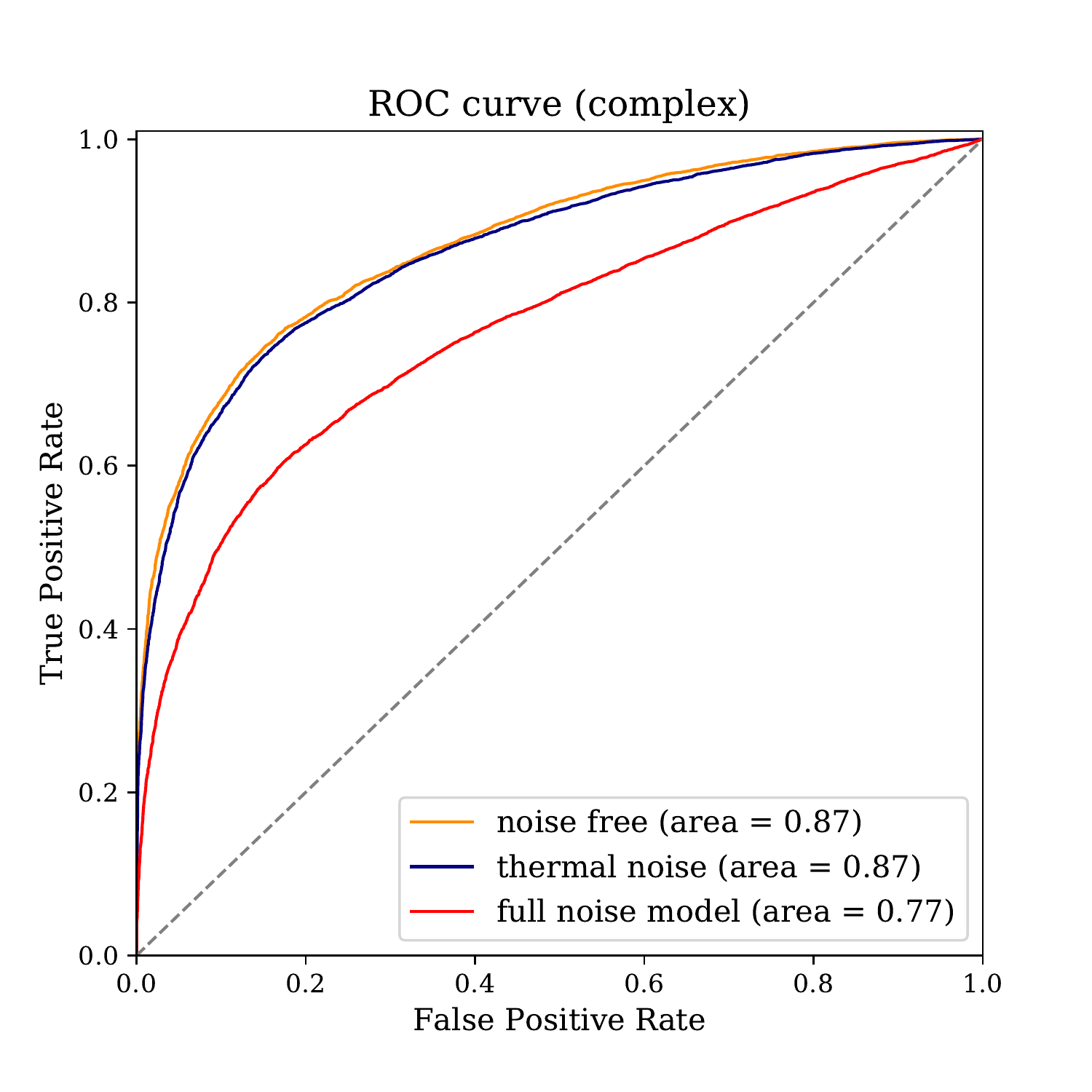}&
\includegraphics[width=0.5\linewidth]{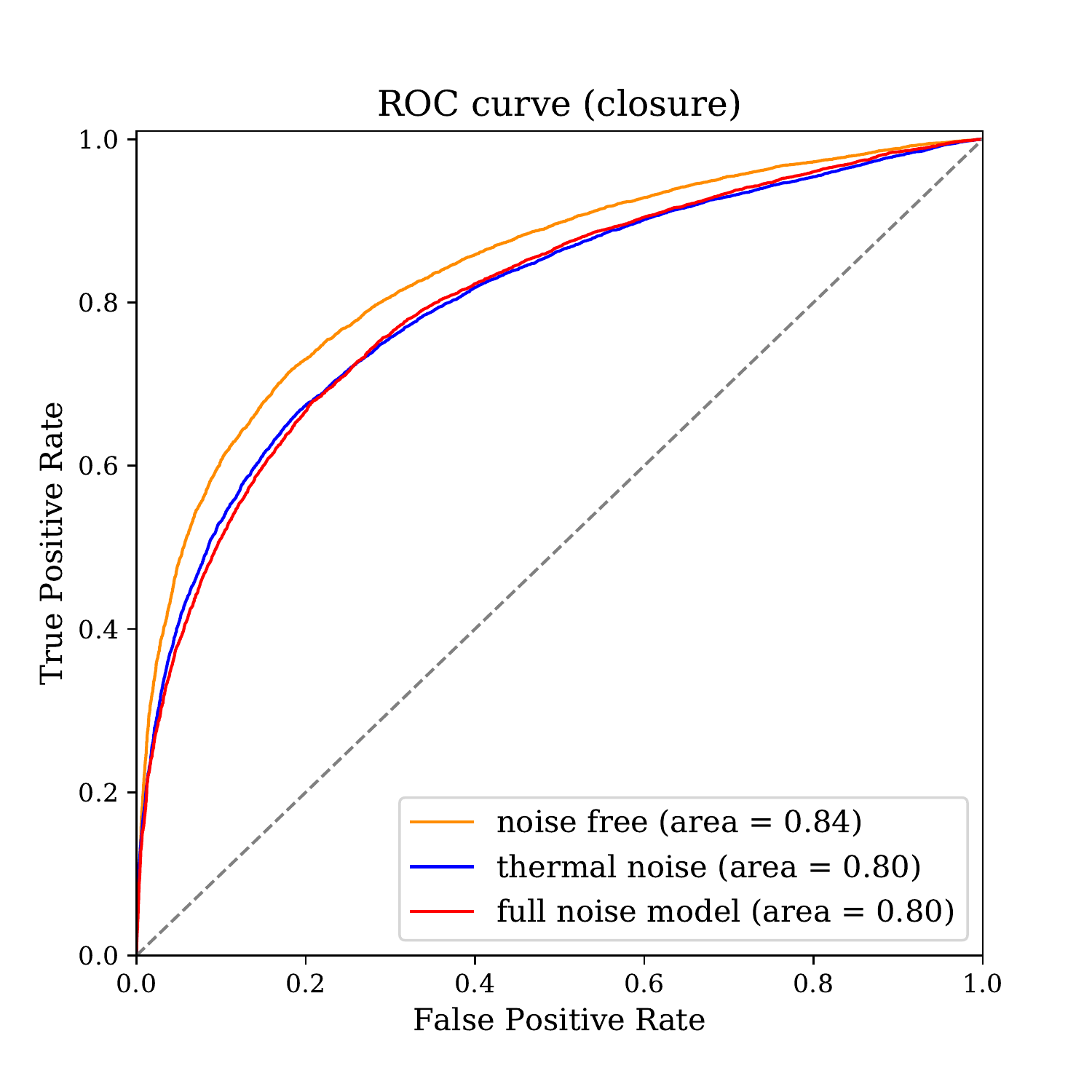}&
\end{tabular}
\caption{
\JL{We calculate the area under Receiver operating characteristic (ROC) curve as a performance measure for different noise models along with complex visibilities/ closure quantities. Here we treat SANE model as positive.}
}
\label{fig:ROC}
\end{figure*}

Figure \ref{fig:histogram} shows the distribution of MAD/SANE scores for 20,000 samples drawn with uniform probability from the test set.  Notice that the vertical axis is logarithmic.  Evidently the data contain information about whether the source is MAD or SANE, and this information can be detected by the NN. \JL{Figure \ref{fig:ROC} shows the performance of our VLBInet with different noise models and the form of VLBI data (complex visibilities/ closure quantities) are given, and we calculate their area under curve (AUC) \citep{bradley1997use}.}

To quantify how well the NN performs in the classification task, we split the data into 20 subsets and calculate the mean and standard deviation of the prediction across the subsets.  The resulting uncertainty estimates are shown in Table \ref{tab:summary_stats}.  

For complex visibilities it is evident that almost no information is lost even when thermal noise is added; the real loss of information is due to gain corruption.  By contrast, the closure quantities lose information via thermal noise but lose nothing through gain corruption.  This is unsurprising since closure quantities are unaffected by gain errors.   

For complex visibility, the MAD/SANE classifier network is able to correctly identify MAD/SANE with $79.67\% \pm 1.13\%$ accuracy without noise, $79.13\% \pm 1.25\%$ accuracy with thermal noise presence and achieve $70.99\% \pm 1.59\%$ accuracy with complex visibility, thermal noise + amplitude gain + phase error. 
For closure visibility, the MAD/SANE classifier network is able to correctly identify MAD/SANE with $76.57\% \pm 1.31\%$ accuracy without noise,  $73.55\% \pm 1.14\%$ accuracy with thermal noise presence and achieve  $73.24\% \pm 1.57\%$ accuracy with complex visibility, thermal noise + amplitude gain + phase error. 



\subsection{Results on real M87* data}

We ran the trained VLBInet (trained on models with added noise and using closure quantities) on real M87 VLBI data taken on  April 5th, 6th, 10th, and 11th, 2017. VLBInet output scores of $0.52$, $0.4$, $0.43$, $0.76$, respectively. Among the 20,000 data in our test set, there are $140$ instances that are similar to the prediction scores (within $0.142$, one sigma from the score distribution for 4 days). Among the $140$ that are similar, $47\%$ of them are ``MAD''.

\begin{figure*}[t]
\vskip 0.2in
\begin{center}
\includegraphics[trim = 2cm 0cm 2cm 0mm,
clip,width=\linewidth]{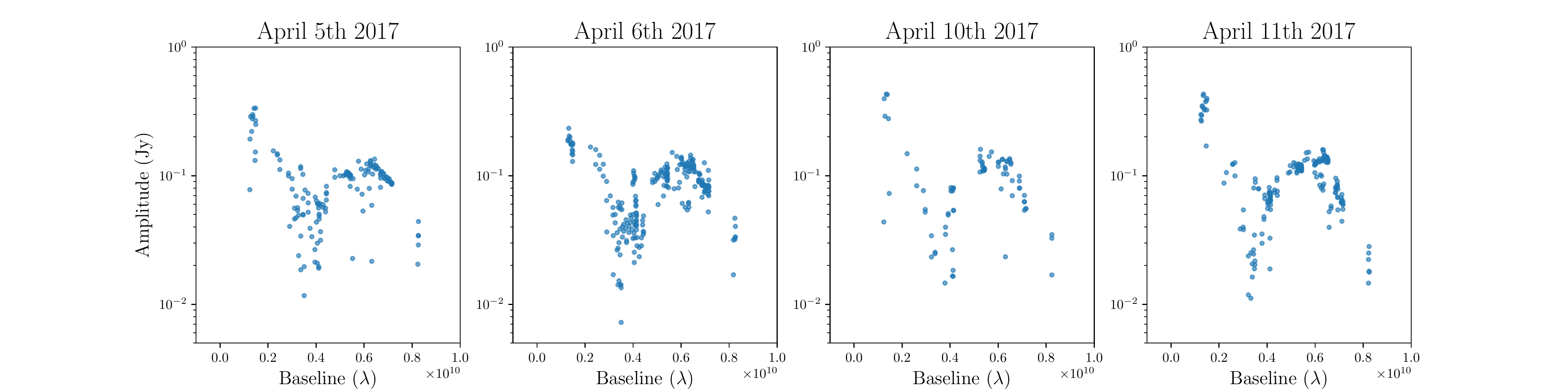}
\caption{Visibility amplitude of M87* taken at 2017 April 5th, 6th, 10th and 11th. The corresponding MAD/SANE scores are $0.52$, $0.4$, $0.43$, $0.76$ for each day's observation.
}
\label{fig:visibility_panel}
\end{center}
\vskip -0.2in
\end{figure*}

It is interesting that the results are similar across days, suggesting that our data are consistent with no strong indication (e.g., over $2 \sigma$) with the MAD/SANE models.  If the models faithfully represent the physical situation in M87*, improvements in data quality (e.g., better baseline coverage) are likely to enable a confident classification of the source as MAD or SANE.   




\begin{table*}
\centering
\begin{tabular}{|p{2cm}||p{4cm}|p{4cm}|p{6cm}|}
 \hline
 \multicolumn{4}{|c|}{Classification accuracy} \\
 \hline
 Visibility form & Noise free & Thermal noise & Thermal noise + complex gains\\
 \hline
 Complex & $79.7\% \pm 1.1\%$ & $79.1\% \pm 1.3\%$ & $71.0\% \pm 1.6\%$ \\
 Closure   & $76.6\% \pm 1.3\%$  &  $73.6\% \pm 1.1\%$ & $73.2\% \pm 1.6\%$\\
 \hline
\end{tabular}

\caption{Summary Statistics \label{tab:summary_stats}}
\end{table*}

\begin{table*}
\centering
\begin{tabular}{|p{4cm}||p{4cm}|p{4cm}|}
 \hline
 \multicolumn{3}{|c|}{Neural Networks MAD/SANE prediction score with closure quantities} \\
 \hline
 Observational date & Score & Prediction \\
 \hline
 2017 April 5 & $0.52$ & SANE \\
 2017 April 6 & $0.40$ & MAD \\
 2017 April 10 & $0.43$ & MAD \\
 2017 April 11 & $0.76$ & SANE \\
 \hline
\end{tabular}

\caption{Neural Networks MAD/SANE prediction score for M87 on all four observational dates  \label{tab:obs_prediction}}
\end{table*}

\section{Discussion}
\label{sec: discussion}

We have presented a neural network pipeline that can perform data classification to identify the accretion state of M87* as either MAD or SANE using VLBI observations, after training on a library of $2 \times 10^4$ simulations.  Our pipeline operates directly on interferometric data products---either complex visibilities or closure quantities---and thus does not rely on an image reconstruction procedure.  We find that the networks are able to achieve a ${>}70$\% classification accuracy across the board, even in the presence of realistic levels of both thermal and complex gain data corruptions.

Our network performs its most accurate classification ($\sim$80\%) when the data are provided as noise-free complex visibilities and its least accurate classification ($\sim$71\%) when the data are provided as complex visibilities containing both thermal and complex gain uncertainties.  The network achieves intermediate performance ($\sim$76\%) when it is provided with noise-free closure quantities and when the closure quantities contain realistic levels of noise the performance is more similar ($\sim$73\%) to that of the corrupted complex visibilities.  These performance differences are qualitatively consistent with the expected relative amounts of source information retained in visibilities versus closure quantities \citep{blackburn2020closure}.  We also note that the classification accuracy using any variant of interferometric data product is considerably worse than that expected when performing classification using the original images, which can achieve essentially perfect ($\sim$99\%; \citealt{lin2020feature}) classification accuracy; again, this difference reflects the loss of information due to sparse Fourier sampling and the reduced angular resolution of the data.

Although we did not provide the neural network with explicit priors, our input training data set covered a limited set of parameter values and thus produced an implicit, data-driven prior through the machine learning process.

\subsection{Comparison with traditional analysis pipelines}

The initial theoretical analysis performed by the EHT included a comparison between observational data products and simulated images of black hole accretion flows.  In a comparison such as this a  $\chi_\nu^2$ (reduced chi squared) distance comparison of real and synthetic data is not useful because intra-model variations---due to changes in image features caused by turbulent fluctuations in the source, for example---are large compared to measurement errors. 

In contrast to snapshot-by-snapshot comparisons, the average image scoring (AIS) procedure described in \citetalias{EHT5} was used to compute the likelihood the data would be drawn from a given model. The AIS procedure ruled out a few retrograde-spin accretion flow models based on variability, but it left much of the parameter space unconstrained. Compared to our data-driven approach, the AIS procedure is less constraining. 


\subsection{Future work}




We have trained VLBInet only to classify M87 source models as MAD or SANE based on the 2017 EHT configuration.  In future work we plan to extend VLBInet to estimate black hole mass and spin as well as parameters describing the state of the plasma, and we plan to consider other arrays such as the EHT 2021 configuration and prospective ngEHT configurations \citep{blackburn2019studying}.  We will also introduce more realistic data corruptions using a more sophisticated end-to-end VLBI synthetic data generation pipeline (e.g., \citealt{roelofs2020symba}).

In this work, we applied the neural network to M87 VLBI data. There are few limitations, however:  1) if our GRMHD and GRRT simulations are flawed then the neural network, which has been trained on a simulation-based synthetic dataset, may not able to generalize to realistic data. 2) Similarly if the noise models used in the generation of synthetic VLBI observables are flawed then the neural network, which has been trained on the flawed noise model in the synthetic dataset, may not be able not able to generalize to realistic noise (e.g., realistic atmospheric, instrumental, and calibration effects) and hence could predict spurious result. 3) Our training set covers the model parameter space sparsely---for example, we use only five values for black hole spin---and the prediction could therefore be biased.

Although we have demonstrated the utility of our framework only for the EHT, we believe that it should in principle work for general classification analyses using interferometric data.  For instance, a similar network trained on an appropriate set of simulated protoplanetary disk observations using ALMA could plausibly constrain properties such as disk inclination, thickness, and mass.  In practice, using the network in such a fashion is likely to be limited by the availability of suitable simulations, perhaps motivating the development of simulation capabilities. This could have broad impact on the community, as interferometric techniques have been crucial for many astrophysical observational cases, including studies of dark matter substructure in strong gravitational lensing \citep{hezaveh2016detection, lin2020hunting} and starburst galaxies \citep{vieira2013dusty}. 

Since the EHT captures information about source polarization and time dependence, we also plan to study whether these additional data can further constrain physical parameters of the source. Synchrotron emission is naturally polarized and the local properties of the magnetic field in the plasma alter the electric vector position angle as radiation propagates through the plasma. Thus, the structure of the magnetic field in the plasma influences the orientation of linear polarization observed in black hole images.
\citet{palumbo2020beta2} and \citetalias{EHT8} showed that the qualitative and quantitative differences between the near-horizon magnetic field strength and structure suggest that linear polarization data might be particular useful in differentiating between MAD and SANE models. We will apply interpretability methods  \citep{ghorbani2019towards, lin2020feature} to gain a more quantitative understanding of the features in the visibility domain. It is critical to understand the origin of accurate neural network classifications/regressions so that we can be confident that we are not overfitting.

As pointed out by \citet{sun2020learning}, the VLBI observing proccess can be viewed as a physics-constrained autoencoder in which the encoder is the sampling strategy through the radio telescope.  Our work could potentially be helpful for evaluating telescope site candidates in a fast and automatic way \citep{raymond2021evaluation}. Visibility in some sense is similar to the hidden representation/bottleneck in the bottleneck of variational autoencoder. Traditionally, one would need to reconstruct the image and do data analysis on top of the reconstructed images. Since the whole VLBI process looks like an auto-encoder, we could in principle not only optimize the decoder (from measurement to parameters) but also optimize the encoder (from source to encoded representation that preserve the most information through sparse sampling) \citep{sun2020deep, bakker2020experimental}.

\acknowledgments

The authors thank He Sun, Katie Bouman, Michael Janssen, Alexander Raymond, Bart Ripperda, Shep Doeleman, Gil Holder, Max Welling, Abhishek Joshi, Vedant Dhruv, David Ruhe, Michael Eickenberg, Shirley Ho, and David Spergel for useful discussion. \JL{The authors thank John Wardle, Shiro Ikeda for useful feedback.}
JL, GNW, BP, and CFG were supported by the National Science Foundation under grants AST 17-16327, OISE 17-43747, and AST 20-34306.  GNW was supported in part by a Donald C.~and F.~Shirley Jones Fellowship and the Institute for Advanced Study.
DWP acknowledges support provided by the NSF through grants AST-1952099, AST-1935980, AST-1828513, and AST-1440254, and by the Gordon and Betty Moore Foundation through grant GBMF-5278. This work has been supported in part by the Black Hole Initiative at Harvard University, which is funded by grants from the John Templeton Foundation and the Gordon and Betty Moore Foundation to Harvard University.
This work used the Extreme Science and Engineering Discovery Environment (XSEDE) resource stampede2 at TACC through allocation TG-AST170024.  This work utilizes resources supported by the National Science Foundation’s Major Research Instrumentation program, grant 1725729, as well as the University of Illinois at Urbana--Champaign. JL thanks the AWS Cloud Credits for Research program. JL and CFG thank the GCP Research Credits program.

\bibliography{sample63}{}
\bibliographystyle{aasjournal}

\end{document}